\documentclass{emulateapj}

\newcommand{\mc}{\multicolumn}
\newcommand{\rxj}{RX~J0420}
\newcommand{\RXJ}{RX~J0420.0$-$5022}
\newcommand{\rxjw}{RX~J1856.5$-$3754}
\newcommand{\rxjk}{RX~J0720.4$-$3125}

\newcommand{\rbsb}{RX~J2143.0+0654}
\newcommand{\rxjvk}{RX~J1605.3+3249}

\newcommand{\xmm}{{\em XMM}}
\defcitealias{kvk05}{KvK05a}
\defcitealias{kvk05b}{KvK05b}
\defcitealias{vkk08}{vKK08}
\defcitealias{kvk09}{KvK09}
\defcitealias{kvk09b}{KvK09b}
\defcitealias{hmz+04}{H+04}
\defcitealias{haberl07}{H07}
\newcommand{\Hzsec}{\ensuremath{{\rm Hz}\,{\rm s}^{-1}}}
\newcommand{\secsec}{\ensuremath{{\rm s}\,{\rm s}^{-1}}}
\newcommand{\expnt}[2]{\ensuremath{#1 \times 10^{#2}}}   % scientific notation

\begin{document}

\shorttitle{The Spin-down of the Nearby Neutron Star \RXJ}
\shortauthors{Kaplan \& van~Kerkwijk}

\title{A Coherent Timing Solution for the Nearby, Thermally Emitting  Isolated Neutron Star
RX~J0420.0$-$5022}
\author{D.~L.~Kaplan\altaffilmark{1} and M.~H.~van Kerkwijk\altaffilmark{2}}

\altaffiltext{1}{Physics Dept., U. of Wisconsin - Milwaukee, Milwaukee
  WI 53211, USA; kaplan@uwm.edu}
\altaffiltext{2}{Department of Astronomy and Astrophysics, University
  of Toronto, 60 St.\ George Street, Toronto, ON M5S 3H8, Canada;
mhvk@astro.utoronto.ca}

\slugcomment{ApJL, in press}
\begin{abstract}
  We present a phase-coherent timing solution for \RXJ, the coolest
  ($kT\simeq45\,$eV) and fastest-spinning ($P=3.45\,$s) of the seven
  so-called isolated neutron stars (INSs).  Using 14 observations with
  the \textit{XMM-Newton} spacecraft in 2010--2011, we were able to
  measure a spin-down rate $\dot \nu=(-2.3\pm0.2)\times 10^{-15}\,{\rm
    Hz\,s}^{-1}$ ($\dot P=\expnt{(2.8\pm0.3)}{-14}\,{\rm s\,s}^{-1}$),
  from which we infer a dipolar magnetic field of
  $\expnt{1.0}{13}\,$G.  With reasonable confidence we were able to
  extend the timing solution back to archival \textit{XMM-Newton} from
  2002 and 2003, giving the same solution but with considerably more
  precision.  This gives \RXJ\ the lowest dipole magnetic field of the
  INSs.  Our spectroscopy does not confirm the broad absorption
  feature at 0.3\,keV hinted at in earlier observations, although
  difficulties in background subtraction near that energy make
  conclusions difficult.  With this, all 6 of the INSs that have
  confirmed periodicities now have constrained spin-downs from
  coherent solutions.  The evidence that the INSs are qualitatively
  different from rotation-powered pulsars now appears robust, with the
  likely conclusion that their characteristic ages are systematically
  older than their true ages, because their fields have decayed.  The
  field decay probably also causes them to be systematically hotter
  than pulsars of the same (true) age.
\end{abstract}

\section{Introduction}
The so-called isolated neutron stars (INSs; see \citealt{haberl07} and
\citealt{kaplan08} for reviews) are a group of seven nearby
($\lesssim\!1\,$kpc) neutron stars with low
($\sim\!10^{32}{\rm\,erg\,s^{-1}}$) X-ray luminosities and long
(3--10\,s) spin periods \citep{kvk09b}.  They are unique in that the
X-ray emission likely comes from a large fraction of the neutron
stars' surfaces, and is not influenced by accretion (as in the case of
X-ray binaries) or non-thermal magnetospheric emission (as in the case
of rotation-powered pulsars); the INSs are also radio-quiet \citep[e.g.,][]{kml+09}.  The
INSs then have the potential to help understand neutron star radii and
cooling via measurements of their emission areas and luminosities, but
this is made difficult by our inability to realistically model the
X-ray, ultra-violet, and optical emission from these objects (e.g.,
\citealt{hkc+07,kkvkh11}).

Recently it was proposed (\citealt{kvk09b}; \citealt*{pmg09};
\citealt{ppm+10}) that the current temperatures and magnetic fields of
the INSs reflect non-thermal, coupled evolution, where the magnetic
field has decayed in strength, heating the neutron-star surface.
Testing this hypothesis is an important step to using the INSs to
constrain the overall cooling history of neutron stars, and through it
probe their inner structure and composition \citep{yp04}.  To
constrain evolutionary models, measurements of the current magnetic
fields and temperatures of the INSs are required.  Similarly, to
understand the broad absorption features seen at energies of
0.2--0.75\,keV in the spectra of almost all INSs
\citep{haberl07,vkk07}, also requires knowledge of the magnetic
fields, as at these field strengths the assumed transition energies
are field-dependent.  Therefore, we have undertaken systematic
measurements of the dipole magnetic fields for the INSs through
phase-coherent X-ray timing using the \textit{Chandra} and
\textit{XMM-Newton} spacecraft
\citep[][hereafter \citetalias{kvk05,kvk05b,vkk08,kvk09,kvk09b}]{kvk05,kvk05b,vkk08,kvk09,kvk09b}.

Here we measure the spin-down of the INS \object[RX
J0420.0-5022]{\RXJ} (hereafter \rxj), the last INS with a confirmed
period measurement but without a timing solution.  \rxj\ was
identified as a possible neutron star by \citet*{hpm99}, although it
had originally been associated with a nearby galaxy.  Followup
\textit{ROSAT}\ observations showed a very soft thermal spectrum and
yielded an improved position, both of which led to its classification
as a neutron star.  While initially a 22.7-s period was suggested,
observations with \xmm\ found a period of 3.45\,s instead, the
shortest among all INSs \citep[][hereafter
\citetalias{hmz+04}]{hmz+04}; the same observations also showed that
\rxj\ was the coolest INS, with $kT\simeq45\,$eV.

\section{Observations \& Analysis}
\label{sec:obs}
We observed \rxj\ fourteen times with \xmm\ \citep{jla+01} in 2010 and
2011, and focus here on the data taken with the European Photon
Imaging Camera (EPIC) with pn and MOS detectors, all used in small
window mode with thin filters (Table~\ref{tab:obs}).  We reprocessed
our observations with SAS version 11.0.0 and calibration files current
as of 2011 May 25.  We also reprocessed the pn data from
\citetalias{hmz+04}, which are taken with the same filter, but with
the full window mode instead (we did not use their full-frame MOS
data, since these do not resolve the pulsations).  We used {\tt
  epchain} and {\tt emchain} and selected source events from a
circular region of $37\farcs5$ radius.  For the pn, we selected
energies between 130 and 800\,eV, where we set our lower energy cutoff
slightly below the default of 150\,eV to increase the net number of
counts from our very soft target (by about 25\%; for even lower
thresholds, the instrumental background increases too rapidly), and
the upper cutoff at the energy at which the source becomes
undetectable (thus minimizing the effects of flares, which dominate
the background at higher energies).  Following standard practice, we
included only one and two-pixel (single and double patterns 0--4)
events with no warning flags for pn, and single, double, and triple
events (patterns 0--12) with the default flag mask for MOS1/2.  We
barycentered the event times using the {\em Chandra X-ray Observatory}
position from \citetalias{hmz+04}: $\alpha=04^{\rm h}20^{\rm
  m}01\fs95$ and $\delta=-50\degr22\arcmin48\farcs1$ (J2000).  We
extracted background lightcurves for pn from similarly sized regions
offset from the source, but at the same \texttt{RAWY} coordinate, as
recommended by the SAS User Guide.\footnote{See
  \url{http://xmm.esac.esa.int/external/xmm\_user\_support/documentation/sas\_usg/USG/node64.html}.}
For MOS1/2, the small-window mode does not permit such large
background areas, but we used several smaller areas to compensate.

\setlength{\tabcolsep}{3pt}
\begin{deluxetable}{c c c c c c}
\tablewidth{0pt}
\tabletypesize{\footnotesize}
\tablecaption{Log of Observations and Times of Arrival\label{tab:obs}}
\tablehead{
&&\colhead{Exp.\tablenotemark{a}}&&\colhead{$f_{\rm bg}$\tablenotemark{a}}& \colhead{TOA\tablenotemark{b}}\\
\colhead{Rev.}&
\colhead{Date}&\colhead{(ks)}&\colhead{Counts\tablenotemark{a}}&\colhead{(\%)}&\colhead{(MJD
  TDB)}\\[-2.2ex]
}
\startdata
% plock[paper]% pyinfo.py -p OBS_ID -p REVOLUT -p DATE-OBS -p TELAPSE ../../xmm/red/*/*PN*SREVLI*.FIT -e 1
%\dataset[ADS/XMM\#0106260201]{\phn168} &2000~Nov~08     & 15.6   & 25,604  &51856.691599(3) \\
\dataset[ADS/XMM\#0141750101]{\phn560} & 2002 Dec 30&	20,047    & 4,201 &10.0&  52638.2855519(12)\\
\dataset[ADS/XMM\#0141751001]{\phn561} & 2002 Dec 31&	20,048    & 4,593 &11.7&  52640.0466236(12)\\
\dataset[ADS/XMM\#0141751101]{\phn570} & 2003 Jan 19&	20,547    & 4,647 &11.6&  52658.8319656(8)\phn \\
\dataset[ADS/XMM\#0141751201]{\phn664} & 2003 Jul 25&	20,036    & 4,384 &11.8&  52846.0226435(10)\\
\dataset[ADS/XMM\#0651470201]{1887}    & 2010 Mar 30&	\phn7,472 & 2,062 &41.2&  55285.5413317(17)\\
\dataset[ADS/XMM\#0651470301]{1890}    & 2010 Apr 04&	\phn9,072 & 2,679 &45.8&  55290.8432112(23)\\
\dataset[ADS/XMM\#0651470401]{1892}    & 2010 Apr 09&	\phn7,772 & 2,049 &35.4&  55295.4037368(11)\\
\dataset[ADS/XMM\#0651470501]{1913}    & 2010 May 21&	\phn5,471 & 1,462 &32.1&  55337.2761247(20)\\
\dataset[ADS/XMM\#0651470601]{1948}    & 2010 Jul 29&	\phn6,472 & 1,626 &37.2&  55406.6382597(30)\\
\dataset[ADS/XMM\#0651470701]{1975}    & 2010 Sep 21&	\phn9,872 & 2,372 &38.2&  55460.4233059(20)\\
\dataset[ADS/XMM\#0651470801]{1981}    & 2010 Oct 02&	11,672    & 2,980 &34.8&  55472.0349984(22)\\
\dataset[ADS/XMM\#0651470901]{1981}    & 2010 Oct 03&	12,972    & 3,064 &38.6&  55472.8839787(11)\\
\dataset[ADS/XMM\#0651471001]{1981}    & 2010 Oct 04&	16,871    & 4,465 &40.2&  55473.3194015(13)\\
\dataset[ADS/XMM\#0651471101]{1983}    & 2010 Oct 06&	10,471    & 2,586 &36.9&  55476.0218532(19)\\
\dataset[ADS/XMM\#0651471201]{2008}    & 2010 Nov 26&	\phn5,471 & 1,339 &35.2&  55526.4316848(19)\\
\dataset[ADS/XMM\#0651471301]{2032}    & 2011 Jan 13&	15,471    & 4,223 &44.5&  55575.0260821(17)\\
\dataset[ADS/XMM\#0651471401]{2071}    & 2011 Mar 31&	\phn7,018 & 1,577 &38.2&  55651.9068660(24)\\
\dataset[ADS/XMM\#0651471501]{2076}    & 2011 Apr 11&	\phn5,471 & 1,248 &32.2&  55662.3338203(19)\\
\enddata
\tablecomments{All observations used  the small window
  mode and  thin filter for  both EPIC-pn and EPIC-MOS1/2,
  except for Revs.~560, 561, 570, 664, in which the full window mode was
  used (which meant that only the EPIC-pn data were suitable for timing).}
\tablenotetext{a}{The exposure time,  number of counts, and estimated
  fraction of events due to background $f_{\rm bg}$ given here
  are for EPIC-pn only.}
\tablenotetext{b}{The TOA is defined as the time of maximum light of
  the fundamental closest to the middle of each observation computed
  from the combined EPIC-pn and EPIC-MOS1/2 datasets, and is given
  with 1-$\sigma$ uncertainties. }
\end{deluxetable}

\subsection{Timing Analysis}
\label{sec:timing}
Our timing analysis largely follows the procedure described in
\citetalias{kvk05}.  As a starting place, we first determined the frequency
that maximized the power in a $Z_1^2$ periodogram for the EPIC-pn data
from the longest observation in Rev.~1981.  We then expanded the
periodogram to include data from all observations in Revs.~1981 and
1983, finding a best-fit frequency of $\nu=0.2896033\pm0.0000003$\,Hz,
consistent with that found by \citetalias{hmz+04} for the earlier
data. In contrast to some of the other INSs, there is no evidence for
higher harmonics in the periodogram: the $Z_1^2$ power is 33.8, while
$Z_2^2=34.7$ and $Z_3^2=35.2$, both of which are consistent with the
additional power
of 1 expected for noise.  We also checked to see if the true period
was in fact 6.9\,s (closer to that of the other INSs), but the pulse
shape, hardness ratio, and median energy were all consistent with no
variation between the first and second halves of the pulse (with 16
bins the lightcurve variation between the first and second halves has
$\chi^2=10.6/8$, while the hardness ratio variation has
$\chi^2=4.1/8$).

\begin{figure}
% timing/plot_soln_pulse.m
%\plotone{timing_pulse.eps}
\plotone{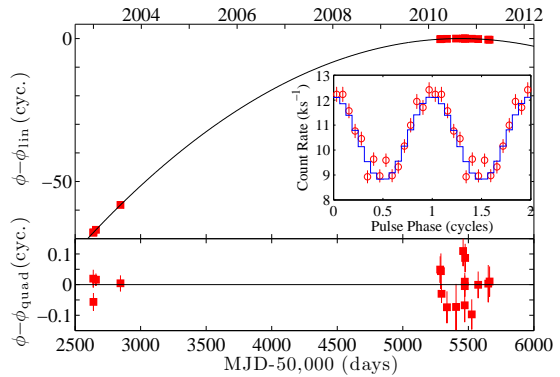}
\caption{Phase residuals for \rxj.  In the top panel, we show the
  residuals relative to a linear model ($\dot\nu=0$).  The line shows
  the best-fit quadratic solution.  Residuals relative to the
  quadratic solution are shown in the bottom panel.  Inset: pulse
  profile for \rxj, based on a combination of all of the 2010--2011
  data folded according to the ephemeris in Tables~\ref{tab:ephem} and
  corrected for the background.  The blue curve is the best-fit
  sinusoidal profile.}
\label{fig:resids}
\end{figure}

Using the above frequency, we determined the times-of-arrival (TOAs;
see Table~\ref{tab:obs}) for the combined EPIC data from each
observation by fitting the binned lightcurves (following
\citetalias{kvk05b}) to a single sinusoid, appropriate given the results
of the periodograms (Fig.~\ref{fig:resids}); the best-fit sinusoid to
the composite, background corrected pn data had a semi-amplitude of
$15\pm2$\%, where the uncertainty includes an estimate for the
variation in the background correction over different background regions.
This is a little higher than the semi-amplitude of about 10\%
from \citetalias{hmz+04}, but differences in background-subtraction
and energy selection could account for the difference (our best-fit to
the 2002--2003 data has a semi-amplitude of $13.2\pm1.1$\%).  The
$\chi^2$ for the fit to the composite profile was good, 14.4 for 13
dof.

Using our TOAs, we were able to identify a reasonably unambiguous
coherent timing solution.  This was possible as we restricted
solutions to have $|\dot \nu|<\expnt{9}{-13}\,\Hzsec$ or $B_{\rm
  dip}<\expnt{2}{14}\,$G (based on the incoherent limits set by
\citealt{haberl07}).  Among those, the solution presented in
Table~\ref{tab:ephem} was the best, yielding $\chi^2=19.9$ for 11
degrees of freedom, with alternatives at $\chi^2=25.1$ ($B_{\rm
  dip}=\expnt{1.8}{14}\,$G), $\chi^2=37.3$ ($B_{\rm
  dip}=\expnt{7}{13}\,$G), and $\chi^2=38.5$ ($\dot \nu>0$).  Of
these, all but the first can be excluded on statistical grounds. The
first comes from an uncertainty of $\pm 1\,$cycle in the cycle count
between the densely sampled Rev.~1981--1983 group and the next closest
observation, Rev.~1975, and is close to the limit from
\citet{haberl07}; we will return to this alias shortly.  We were able
to identify the same solution using a single coherent $Z_1^2(\nu,\dot
\nu)$ periodogram (as in \citetalias{vkk08}).  Spin-down is
well-detected, at $\sim\!10\,\sigma$.  The reduced $\chi^2$ is
somewhat high, but even adjusting our uncertainties to allow for an
reduced $\chi^2$ of 1 will still give an $8\sigma$ detection of
spin-down.  The implied magnetic field is well within the range of
other detections for the INSs \citepalias{kvk09b}.

We can confirm and improve our solution by extrapolating it back to
the older data from 2002--2003.  The time difference is roughly
2600\,days and our $\dot \nu$ uncertainty gives a formal cycle-count
uncertainty of $\pm5\,$cycles, but by trying multiple solutions, we
find that only a single cycle count difference leads to a solution
that fits all four earlier TOAs.  Trying this generally, iteratively
exploring all cycle-count ambiguities between all data sets, we find a
single best-fit solution that agrees with the best-fit solution using
only the new data (Table~\ref{tab:ephem} and Figure~\ref{fig:resids}).
With $\chi^2=24.8$ for 15 degrees-of-freedom, it still has a slightly
high reduced $\chi^2$.  However, the alternate solutions either have
nearly the same implied spin-down rate but differ slightly in the
cycle counts between the 2002--2003 and 2010--2011 observations (and
the lowest of those has $\chi^2=33.0$), or are dramatically different
but have significantly worse $\chi^2$ ($\chi^2=45.9$ and $\dot
\nu>0$).  The alternate, $B_{\rm dip}=\expnt{1.8}{14}\,$G solution
from above does not extrapolate well to the older data, allowing us to
exclude it.  Overall, the older data thus allow us to confidently
select the correct solution to the new data, and select a reasonably
secure overall solution ($\Delta \chi^2=8.1$).

\begin{deluxetable}{c c c}
\tablewidth{0pt}
%\tabletypesize{\footnotesize}
\tablecaption{Measured and Derived Timing Parameters for \RXJ\label{tab:ephem}}
\tablehead{
\colhead{Quantity} & \mc{2}{c}{Value} \\
 & \colhead{2010--2011} & \colhead{2002--2011}\\
}
\startdata
Dates (MJD) \dotfill & 55,286--55,662 & 52,638--55,662\\
$t_{0}$ (MJD)\dotfill        &55,430.6001387(6) & 55430.6001387(6)\\
$\nu$ (Hz) \dotfill          &0.2896029061(12) & 0.2896029058(10)\\
$\dot \nu$ ($10^{-15}\,$\Hzsec)&$-2.3(2)$ & $-2.314(8)$\\
TOA rms (s) \dotfill         & 0.3  & 0.2\\
$\chi^2$/DOF \dotfill        & 19.9/11 & 24.8/15\\
$P$ (s)\dotfill              & 3.453004024(14) & 3.453004027(12)\\
$\dot P$ ($10^{-14}\,$\secsec)\dotfill   & 2.8(3) & 2.759(10)\\
$\tau_{\rm char}$ (Myr)\dotfill& 2.0 & 2.0\\
$B_{\rm dip}$ ($10^{13}\,$G) \dotfill   &1.0 & 1.0\\
$\dot E$ ($10^{31}\,{\rm erg}\,{\rm s}^{-1}$)\dotfill&2.7 & 2.7\\
\enddata
\tablecomments{Quantities in parentheses are the formal 1-$\sigma$
  uncertainties on the last digit.  $\tau_{\rm char}=P/2{\dot P}$ is
  the characteristic age, assuming an initial spin period $P_0\ll P$
  and a constant magnetic field; $B_{\rm
    dip}=\expnt{3.2}{19}\sqrt{P{\dot P}}{\rm\,G}$ is the magnetic field
  inferred assuming spin-down by dipole radiation; $\dot
  E=\expnt{3.9}{46}\nu\dot\nu\,{\rm erg\,s^{-1}}$ is the spin-down luminosity.  
 }
\end{deluxetable}

\subsection{Spectroscopic Analysis}
\label{sec:spectra}
We examined all EPIC-pn spectra of \rxj.  (A full spectral analysis,
including the EPIC-MOS and RGS data and a phase-resolved analysis, is
in progress.)  We used the same source and background extraction
regions as for the timing analysis, created appropriate response
files, and binned the spectral files such that the number of source
plus background counts was at least 25 and the bin width was at least
30\,eV (so that there are roughly 2 bins per EPIC-pn resolution element).

\begin{figure}
%spectrum/plot_mergedfits.m
%\plotone{merged_bbfit.eps}
\plotone{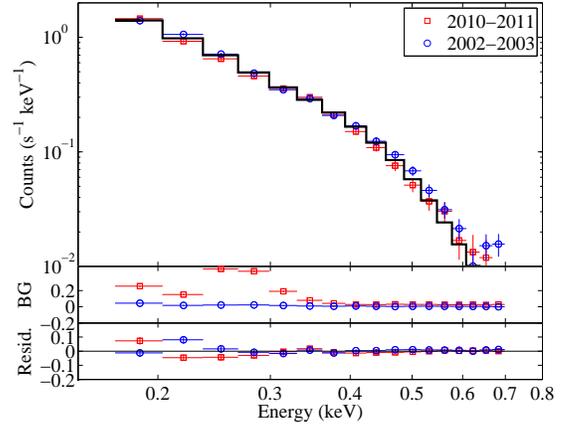}
\caption{Blackbody fits to the merged EPIC-pn data of \rxj, from
  2002--2003 (blue circles) and 2010--2011 (red squares).  The model
  is the thick black line.  The middle panel shows the backgrounds
  used, and highlights the background structure seen at low energies
  in the 2010--2011 data.  The bottom panel shows the residuals.}
\label{fig:bbfit}
\end{figure}

We first compared the raw EPIC-pn spectra of all of the observations
against each other.  This did not include any response files or
calibration corrections, but even so the binned pn spectra were
generally consistent with each other, implying no spectral change
(Fig.~\ref{fig:bbfit}).  There are small deviations at low energies
(0.2--0.4\,keV) and we will return to these below, but the 2010-2011
data did not show any appreciable variability.

We fit the pn data using \texttt{sherpa} \citep{sherpa}.  To aid in
fitting we merged the event and response files into two groups: one
for 2002-2003 (full-frame data, also fit by \citetalias{hmz+04}) and
one for 2010-2011 (small-window data).  While not perfect, we found
that an absorbed blackbody provided a reasonable fit, with $N_{\rm
  H}<\expnt{1}{18}\,{\rm cm}^{-2}$, $kT_\infty=47.6\pm0.3\,$eV, and
$R_{\infty}=12.8\pm0.3\,{\rm km\,kpc}^{-1}$ (formal 1-$\sigma$
uncertainties; $\chi^2=73.0$ for 31 dof), reasonably consistent with
\citetalias{hmz+04}.  Unlike \citetalias{hmz+04}, we do not find
evidence for a spectral feature, perhaps because of changes in the
response files and calibration since earlier fits.  However, there are
indications that our fit is not completely reliable.  First, the
best-fit value of the absorption is 0 (although it is covariant with
the blackbody temperature).  Second, there are some residuals near
0.33\,keV, where \citetalias{hmz+04} found evidence for a spectral
line.  However, we could not find a consistent fit to both sets of
data.  Third, stronger differences are seen at energies of
0.2--0.3\,keV (see Figure~\ref{fig:bbfit}).  Some of these are too
narrow to come from astrophysical sources, and instead likely reflect
problems in background subtraction (the low-energy background for the
small-window data in particular can be significant and has substantial
energy structure\footnote{See
  \url{http://www.star.le.ac.uk/$\sim$amr30/BG/BGTable.html}.}).  Hence,
it is difficult to interpret any residual structure there with
respect to a blackbody.

\section{Discussion \& Conclusions}
We have determined a reliable, statistically significant coherent
spin-down solution for \rxj.  With this, only \rxjvk, which as yet has
only a tentative detection of a periodicity, lacks a coherent solution
(although in the cases of \rbsb\ and RX~J0806.4$-$4123 spin-down was
not well measured, and further observations are in progress).  While
the overall results of our timing program have been discussed at
length in previous papers (overall energetics in \citetalias{vkk08};
spectral implications in \citetalias{kvk09}; evolutionary models in
\citetalias{kvk09b}), here we touch on some of the aspects that make
\rxj\ unique and compare it to the other objects in its class.

First, while the timing properties (dipole magnetic field, $\dot E$,
characteristic age) of \rxj\ place it well within the INSs, \rxj\ has
the shortest period by more than a factor of~2.  In the context of the
magneto-thermal evolution model, this could be a consequence of a
lower initial magnetic field, and thus less dramatic early spin-down.
\rxj\ also has the lowest current field, although it is not clear
whether there is a good correlation between current magnetic field and
period: RX~J0806.4$-$4123 and RX~J2143.0+0654 both have long periods
but relatively weak magnetic fields.  The magnetic field of \rxj\ is
low enough that it would not be remarkable in a radio pulsar.

Second, the temperature of \rxj\ is the lowest of the INSs.  Again,
this might make sense if it is roughly the same age as the rest of the
INSs but started with the lowest magnetic field.  It would then have
been heated the least, and would come closest to the ``pristine''
cooling of a non-magnetic neutron star.  In this context, it is
interesting that \rxjw\ is cooler and less magnetized than \rxjk\ (1.5
vs $2.5\times10^{13}\,$G), while appearing the younger one by
kinematic age (\citealt*{kvka07}; \citealt{tnhm10,tenh11}).  It would
be interesting to measure the kinematic age of \rxj\ to compare it
with the rest of the population.  

\begin{figure}
%\plotone{kT_tau_log.eps}
\plotone{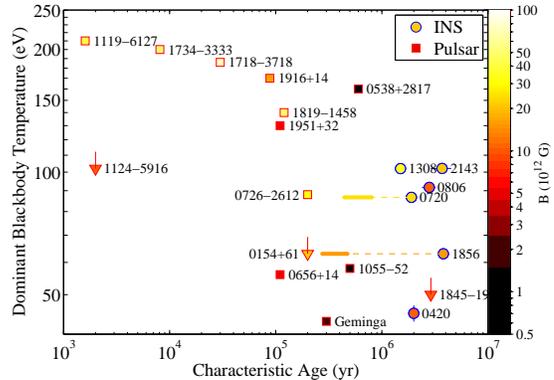}
\caption{Blackbody temperature (measured at infinity) versus
  characteristic age for the INSs (circles) and rotation-powered
  pulsars (squares and upper limits); for all objects the color
  indicates the dipole magnetic field according to the scale at the
  right.  The pulsars included here are both those that are visible in
  the \textit{ROSAT} All-Sky Survey and those with dipole fields $\geq
  10^{13}\,$G that have X-ray measurements.  See \citet{zkm+11} for a
  similar plot, although there they emphasize the variation among
  radio pulsars as a function of magnetic field, while here we
  emphasize primarily the distinction between pulsars and INSs.  The
  data are taken from \citet{kvk09b}, with the addition of
  PSR~B1916+14 \citep{zkg+09}, PSR~J1734$-$3333 \citep{oklk10}, an
  updated temperature for PSR~J1718$-$3718 \citep{zkm+11}, an upper
  limit for PSR~B1845$-$19 (Zhu et al.\ 2011, in prep),
  PSR~J0726$-$2612 (Kaplan et al., in prep.), an updated timing
  solution for RX~J0806.4$-$4123 (van Kerkwijk et al., in prep.), and
  data from this paper.  We only show data for sources where the
  blackbody can plausibly represent the majority of the surface (i.e.,
  $R_{\rm BB}>3\,$km, allowing for distance uncertainties).  This is
  not meant to be a cooling curve, but instead to demonstrate the
  inconsistency in characteristic ages for the INS relative to the
  pulsars at the same $kT$.  The horizontal lines extending to the
  left of \rxjw\ and \rxjk\ show the ranges of plausible kinematic
  ages \citep{kvka07,tenh11} for those objects, with the thick parts
  the mostly likely ranges.}
\label{fig:kTtau}
\end{figure}

To view the evidence for field decay in a different way, we show in
Figure~\ref{fig:kTtau} the blackbody temperature versus characteristic
age for pulsars and the INSs (see also \citealt{zkm+11}).  It is quite
clear that the INSs are systematically a factor of 5--10 older in
characteristic age for the same temperature.  If instead one uses the
kinematic age, however, one sees that the difference is much smaller
(for the two sources for which kinematic ages are available).  In the
context of a picture in which the fields of INSs decayed, the main
difference with pulsars induced by the initially much stronger field
is thus that it leads to rapid initial spin down and long present
periods (and thus long characteristic ages); the current temperatures
are not as strongly affected.  Indeed, in the models of \citet{pmg09},
the heat generated by field decay is lost fairly rapidly.

Third, given both the low temperature and low magnetic field, \rxj\ largely
follows the empirical temperature-magnetic field correlation from
\citetalias{kvk09}. As discussed there, the origin of this relation
(evolutionary vs.\ surface physics) or even its overall integrity in the
face of new data are not clear.  It does seem to form an upper limit
to the possible magnetic field of an INS, and even the rotating radio
transient (RRAT) J1819$-$1458 (possibly somewhat younger than the
INSs, and with higher $\dot E$ as well) seems to roughly agree
\citep[based on][]{mrg+07}.

Fourth, we did not confirm the tentative absorption feature found by
\citetalias{hmz+04}, although problems with the background subtraction
meant we cannot refute it with confidence either.
If it is true that \rxj\ has no broad X-ray absorption feature, it
would join \rxjw\ (although as this object is often used for calibration
{\em assuming} it emits like a blackbody, it is difficult to set
confident limits).  These are the two INSs with the lowest
temperatures and the lowest magnetic fields, suggesting some relation
between the presence of absorption features (or their energy) and
either temperature or field strength (although a direct correlation of
energy with field strength seems excluded; \citetalias{kvk09}).
\rxj\ is also similar to \rxjw\ in its optical excess: both are
reasonably well fit by Rayleigh-Jeans like powerlaws, unlike the other
INSs whose spectra are softer \citep{kkvkh11}.  It is possible that
the optical/UV spectral index is related to the magnetic field, either
directly through the magnetosphere \citep{txs11} or indirectly through
shifting spectral lines \citep{kkvkh11}; in that case the similarity
of \rxj\ and \rxjw\ would be natural.

Overall, our measurement firmly places \rxj\ as one of the INSs
despite its short period, and moves us significantly closer to having
a complete sampled of measured spin-downs for that population.  There
are still a number of open questions to be answered via X-ray and
multi-wavelength observations.  Primary among these is understanding
the surface emission through consistent modeling of the spectra and
lightcurves, and ideally with phase-resolved spectroscopy.  
Observations at optical/UV wavelengths of the pulsed emission could
make significant improvements in our understanding, by tying the
emitting areas at different wavelengths together and establishing the
degree of surface inhomogeneiety.  Finally, further kinematic ages
would help greatly in constraining the coupled evolution of magnetic
field and temperature.

\acknowledgements Based on observations obtained with XMM-Newton, an
ESA science mission with instruments and contributions directly funded
by ESA Member States and NASA.  DLK was partially supported by NASA
through grant NNX08AX39G.  Apart from the XMMSAS data reduction
pipelines provided by \textit{XMM-Newton}, this research has made use
of software provided by the Chandra X-ray Center (CXC) in the
application packages CIAO and Sherpa.

%\bibliography{ins}

%% --------------------------------------------------------------------
%% Thu Aug  4 17:02:29 2011
%%   This file was generated automagically from the files
%%   ms.bbl and ms.tex using
%%     /Users/dlk//perl/nat2jour.pl
%%   This file should accompany ms-aas.tex.
%% --------------------------------------------------------------------

\end{document}